\documentstyle[epsf]{article}

\def\abstract#1{\vskip 7mm 
        \begin{center}{\large Abstract}\par \smallskip
                \begin{minipage}[c]{12cm}
                        \small #1
                \end{minipage}
        \end{center}
}
\def\title#1{\begin{center}{\Large\bf #1}\end{center}}
\def\author#1{\vskip 5mm \begin{center}{#1}\end{center}}
\def\address#1{\begin{center}{\it #1}\end{center}}

\renewcommand{\cite}[1]{\ref{#1}}

\newcommand{\half}{\frac{1}{2}}
\newcommand{\reflef}{(\ref}
\newcommand{\beq}{\begin{equation}}
\newcommand{\eeq}{\end{equation}}
\newcommand{\beqa}{\begin{eqnarray}}
\newcommand{\eeqa}{\end{eqnarray}}
\newcommand{\bcent}{\begin{center}}
\newcommand{\eeeqa}{\end{eqnarray*}}
\newcommand{\bbeqa}{\begin{eqnarray*}}
\newcommand{\ecent}{\end{center}}
\newcommand{\bitem}{\begin{itemize}}
\newcommand{\eitem}{\end{itemize}}
\newcommand{\bdesc}{\begin{description}}
\newcommand{\edesc}{\end{description}}
\newcommand{\bminip}{\begin{minipage}}
\newcommand{\eminip}{\end{minipage}}
\newcommand{\barray}{\begin{array}}
\newcommand{\earray}{\end{array}}

\makeatletter
\@ifundefined{lesssim}{}{}
\@ifundefined{gtrsim}{}{}
\def\vereq#1#2{\lower3pt\vbox{\baselineskip1.5pt \lineskip1.5pt
\ialign{$\m@th#1\hfill##\hfil$\crcr#2\crcr\sim\crcr}}}
\makeatother

\begin{document}

\title{%
  Cosmological Constant, Quintessence and Scalar-Tensor Theories of
Gravity\footnote{Delivered at The Ninth Workshop on General Relativity
and Gravitation, Hiroshima University, Nov. 3--6, 1999: International
Seminar Nara 99, Space, Time and Interactions, Nara Women's University, December 4--5, 1999. To appear in Proceedings.}
  \smallskip \\
}
\author{%
  Yasunori Fujii,\footnote{E-mail: fujii@handy.n-fukushi.ac.jp}
}
\address{%
Nihon Fukushi University,
Handa, 475-0012, Japan
}

\abstract{
We show how the scalar field, a candidate of quintessence, in a proposed model of the scalar-tensor theories of gravity provides a way to understand a small but nonzero cosmological constant as indicated by recent observations.  A particular emphasis is placed on the effort to inherit the success of the scenario of a decaying cosmological constant.  Discussions of a possible link to non-Newtonian gravity, the coincidence problem, the issue of time-variability of coupling constants as well as the chaos-like nature of the solution  are also included in a new perspective.
}

\section{Introduction}

For the last few years we have heard much about indications, coming from such observations like age problem, large-scale structure, lens statistics and type Ia supernovae, for the presence of a small but nonzero cosmological constant.  The expected size is measured in units of the parameter $\Omega_{\Lambda}\equiv \Lambda /\rho_{\rm crit}$, with the most favored number somewhere around $0.6 - 0.7$ [\cite{perl}].

Acceptance of the conclusion from a theoretical point of view is limited, however, because $\Lambda$ has been considered to be highly artificial, unnatural and arbitrary.   Alternative ideas have been discussed to produce nearly the same end result.  One of the key observations is that in the language of ideal fluid in cosmology, introducing $\Lambda$ is equivalent to assuming an equation of state, $p=-\rho$.  This aspect of negative pressure is what one can implement in terms of a scalar field that couples to ordinary matter roughly as weakly as gravity.

The study of this type of a ``gravitational scalar field" has a long history [\cite{jordan}--\cite{peeb}].  Recently, Chiba {\sl et al}. presented a detailed analysis showing how the assumed negative pressure can be compared with observations [\cite{chiba}].  They did not confine themselves to a scalar field, calling it X-matter.  Caldwell {\sl et al}. were the first to use a new name ``quintessence," as the fifth element of the universe in addition to the four; baryons, radiation, dark matter and neutrinos [\cite{stein}].  They focussed more on the scalar field, discussing what the potential of the field should be like in order to fit the observations.  An inverse-power potential is preferred over the exponential potential, for example.

We complain, however, that many of the analyses are too much phenomenology-oriented [\cite{qt},\cite{qt2}].  We are talking about $\Lambda$ which has been much discussed, often under such a grandeur name like the ``cosmological constant problem," at the center of which lies a question why the observed value or its upper bound is so small compared with what we expect naturally from a theoretical point of view, by as much as 120 orders of magnitude [\cite{wnb}].

In this connection we remind the audience that some progress has been
achieved in terms of scalar-tensor theories.  We believe that these
theories are promising because they might have important aspects
shared by attempts toward unified theories, and also because they
provide the simplest derivation of an exponential potential of the
scalar field, allowing us to implement the ``scenario of a decaying
cosmological constant" [\cite{decay}].  According to this scenario, we find $\Lambda$
not to be a true constant, but to decay like $\Lambda (t)\sim t^{-2}$.
In the (reduced) Planckian unit system of $c = \hbar = 8\pi G =M_{\rm P}^{-2}=1$, the present age of the universe, $t_0 \sim 10^{10}{\rm y}$ is about
$10^{60}$, hence giving $\Lambda (t_0)\sim 10^{-120}$; today's
cosmological constant is small only because our universe is old, not
because of an unnatural fine-tuning of the parameters to the accuracy 
of 120 orders of magnitude.

This ``success,'' however, is not good enough for understanding the
accelerated universe as suggested by the recent observations, which
indicates $\Lambda (t)$ that behaves nearly constant for some duration 
of time toward the present epoch, or at least falling off more slowly
than $\sim t^{-2}$.  This requires a deviation from the purely
exponential potential, though without losing its advantage of
implementing the scenario of a decaying $\Lambda (t)$.  Proposed
modification of the prototype Brans-Dicke (BD) model will be presented 
in Section V, which will be preceded, however, by Sections II--IV.  In Section II, we show that the prototype BD model with $\Lambda$ added entails a serious drawback.  A remedy is proposed in Section III.  A possible link to non-Newtonian gravity is suggested in Section IV.  Section VI is for discussion including the issue of time-variation of coupling constants, the novel view on the coincidence problem, the expected chaos-like nature of the solution, and possible extensions of the model.

The present article is based on the talks intended to outline basic ideas and main conclusions of Refs. [\cite{yfptp}-\cite{yfnn}], in which more details and references will be found.

\section{Prototype BD model with $\Lambda$}

We start with the Lagrangian [\cite{yfptp}]
\beq
{\cal L}_{\rm BD}=\sqrt{-g}\left( \half\xi\phi^2 R -\epsilon\half
g^{\mu\nu}\partial_{\mu}\phi \partial_{\nu}\phi -\Lambda +L_{\rm
matter} \right),
\label{stu3_1}
\eeq
where we use symbols similar to but slightly different from the
original ones [\cite{bd}];
\[
4\xi\omega = \epsilon = {\rm Sign}(\omega),\quad \varphi_{\rm BD} =(\xi /2)\phi^2.
\]
We use the reduced Planckian unit system, in which $\Lambda$ is naturally expected to be of the order unity, as suggested in many of theoretical models of unification.

An important assumption is made that the scalar field is decoupled from 
ordinary matter fields at the level of the Lagrangian.  This feature distinguishes the prototype BD model [\cite{bd}] from the preceding model due to Jordan [\cite{jordan}].  In the field equation, however, we find that the scalar field does couple to the trace of the matter energy-momentum tensor.  This is due to the mixing interaction between the scalar field and the spinless component of the metric tensor implied by the nonminimal coupling term.  The scalar field couples to matter only via the metric field.  This explains why the scalar field in this model respects Weak Equivalence Principle (WEP) which is a privilege of the force endowed with spacetime geometry.

Conformal transformations defined by
\beq
g_{\mu\nu} \rightarrow g_{*\mu\nu} = \Omega^2(x)g_{\mu\nu},
\label{stu3_2}
\eeq
always provide a powerful tool in analyzing theories with a nonminimal coupling, like the first term on the right-hand side of \reflef{stu3_1}).  Of particular importance is the choice
\beq
\Omega = \xi^{1/2}\phi,
\label{stu3_3}
\eeq
by which we eliminate the nonminimal coupling.  In fact the same Lagrangian \reflef{stu3_1}) can be put into the form
\beq
{\cal L}_{\rm BD}=\sqrt{-g_*}\left( \half R_* -\half
g_{*}^{\mu\nu}\partial_{\mu}\sigma\partial_{\nu}\sigma - V(\sigma)
+L_{*\rm matter} \right),
\label{stu3_4}
\eeq
in which the term of the curvature scalar is the standard Einstein-Hilbert (EH) term, and the {\em canonical} scalar field is now $\sigma$ related to the original $\phi$ by
\beq
\phi = \xi^{-1/2}e^{\zeta\sigma},
\label{stu3_5}
\eeq
where
\beq
\zeta^{-2}=6+\epsilon\xi^{-1}.
\label{stu3_6}
\eeq
We assume that the right-hand side is positive;
\beq
\epsilon\xi^{-1} > -6.
\label{stu3_7}
\eeq
Notice also that the term of $\Lambda$ acts now as an exponential potential as alluded before;
\beq
V(\sigma) = \Lambda \Omega^{-4}= \Lambda e^{-4\zeta\sigma}.
\label{stu3_8}
\eeq

We say that we have moved from the original Jordan conformal frame, abbreviated by J frame, to Einstein (E) frame.  Generally speaking, quantities in different conformal frames (CFs) are related to each other in an unambiguous manner.  For this reason different CFs are sometimes called to be equivalent to each other.  Physics may look different, however.  Then a question may arise how we can single out a particular CF out of (infinitely many) CFs.  We do it according to what clock or meter stick we use.  Remember that \reflef{stu3_2}) represents a local change of units [\cite{dpr}], as will be evident if it is put into the form $ds_*^2 = \Omega^2 (x)ds^2 $.  These features will be demonstrated later by explicit examples.

Let us choose E frame for convenience of mathematical analysis, for the moment.  Consider spatially flat Robertson-Walker cosmology.  Looking for the solution in which the scalar field is spatially uniform, we find analytic solution;
\beqa
a_*(t_*)&=&t_*^{1/2}, \label{stu3_9}\\
\sigma (t_*)&=&\bar{\sigma} + \zeta^{-1}\half \ln t_*,
\label{stu3_10} \\
\rho_{m*}(t_*)&=&\left( 1-\frac{1}{4}\zeta^{-2} \right) t_*^{-2}\times \left\{\begin{array}{l} \frac{3}{4}, \\[.5em]
       1,  
\end{array}\right.
\label{stu3_11} \\[.8em]
\rho_{\sigma}(t_*)&=& t_*^{-2}\times \left\{
\begin{array}{l}
\frac{3}{16}\zeta^{-2}, \\[.5em]
\frac{1}{4}\left( -1+\zeta^{-2} \right),
\end{array}
\right.
\label{stu3_12}
\eeqa
where $t_*$ is the cosmic time in E frame, $\rho_{m*}$ is the ordinary matter density, while $\rho_{\sigma} \equiv \dot{\sigma}^2 /2 + V(\sigma)$ may be interpreted as an effective cosmological constant $\Lambda_{\rm eff}$.  The upper and lower lines in \reflef{stu3_11}) and \reflef{stu3_12}) are for the radiation- and dust-dominated universe, respectively.  We first notice that \reflef{stu3_12}) shows that the $\Lambda_{\rm eff}$ decays like $t_*^{-2}$, thus implementing the scenario of a decaying cosmological constant.

The solution \reflef{stu3_9})-\reflef{stu3_12}), which is in fact an attractor reached asymptotically, differs in many ways from the one obtained from the equation without $\Lambda$.  For example, \reflef{stu3_9}) for the scale factor holds true not only for radiation-dominance but also for dust-dominance; the exponent for dust dominance without $\Lambda$ is given by $2/(3+2\zeta^2)$ which is close to $2/3$ if $\zeta^2 \ll 1$.

More important is, however, that particle masses depend on $\sigma$, and hence on time.  The mass term of the quark, for example, is given by
\beq
{\cal L}_{mq} = -\sqrt{-g}m_0\bar{q}q,
\label{stu3_13}
\eeq
with a {\em constant} $m_0$, due to the assumed absence of $\phi$ in the matter Lagrangian.  In E frame, however, we have
\beq
{\cal L}_{mq} = -\sqrt{-g_*}m_*\bar{q}_*q_*,
\label{stu3_14}
\eeq
with $q_* = \Omega^{-3/2}q$, while
\beq
m_* =m_0 \Omega^{-1}=m_0 e^{-\zeta\sigma}\sim t_*^{-1/2},
\label{stu3_15}
\eeq
where we have used \reflef{stu3_10}) to derive the last expression in \reflef{stu3_15}).

If we consider the period of primordial nucleosynthesis, for example, \reflef{stu3_15}) would imply the decrease of nucleon mass, say, as much as $\sim 70$\%.  This is unacceptable because the physical processes are analyzed in terms of quantum mechanics in which particle masses are taken as constant.  The physical CF corresponding to this situation is J frame.  We should then study the solution in J frame.

We may restart with the cosmological equations in J frame.  There is, however, another interesting approach by using the relations
\beqa
dt_* &=&\Omega dt, \label{stu3_16} \\
a_* &=&\Omega a   \label{stu3_17}
\eeqa
which can be obtained directly from \reflef{stu3_2}).  Combining $\Omega \sim \phi \sim t_*^{1/2}$ obtained from \reflef{stu3_3}), \reflef{stu3_5}) and \reflef{stu3_10}), we derive
\beq
a = \mbox{const},
\label{stu3_18}
\eeq
implying that the universe looks {\em static} in J frame.  This can be understood even in E frame by noting that $a_*$ and $m_*^{-1}$ grow in the same way $\sim t_*^{1/2}$, and that assuming time-independent mass is equivalent to choosing the inverse mass as the time unit.

We have a dilemma; varying mass in E frame and static scale factor in J frame.  We admit that the conclusion is not final, because we assumed the simplest nonminimal coupling term and a purely constant $\Lambda$ in J frame.  These can be modified.  More detailed study shows, however, that we must accept rather extreme  choices of parameters.  We would find it increasingly difficult and unnatural to understand why the simple standard cosmology works so well.  It seems better to modify the model at the more fundamental level, still based on simple physical principles.

\section{Dilaton model}

In view of \reflef{stu3_9}) which happens to coincide with the standard result for the radiation-dominated universe, we try to see if we can have a constant mass in E frame instead of J frame.  Let us assume the E frame mass term,
\beq
{\cal L}_{mq} = -\sqrt{-g_*}m_{q\dagger}\bar{q}_*q_*,
\label{stu3_19}
\eeq
where $m_{q\dagger}$ is chosen to be constant.  This can be transformed back to J frame;
\beq
{\cal L}_{mq} = -\sqrt{-g}\xi^{1/2}m_{q\dagger}\bar{q}q\phi,
\label{stu3_20}
\eeq
a simple Yukawa interaction, with the coupling constant $\xi^{1/2}m_{q\dagger}$ which is in fact dimensionless as will be confirmed by re-installing $M_{\rm P}^{-1}$.  We have no mass term in the starting Lagrangian.  The nonminimal coupling implies that we have no dimensional gravitational constant either.  For this reason we have scale invariance, or dilatation symmetry, which might be chosen as a simple physical principle we mentioned above.  We have a conserved dilatation current in J frame.  The presence of mass term \reflef{stu3_19}) and the EH term in E frame suggests that dilatation symmetry is {\em broken spontaneously} with $\sigma$ a massless Nambu-Goldstone boson, called a {\em dilaton}.

Allowing $\phi$ to be present in the J frame Lagrangian implies that WEP is violated.  We still maintain Equivalence Principle at the more fundamental level as stated that tangential space of curved spacetime be Minkowskian.  Breaking WEP as a {\em phenomenological} law can be tolerated within the limit allowed by the fifth-force type experiments [\cite{fsb}].  We have now E frame as a physical CF in which we develop radiation-dominated cosmology showing standard result with constant particle masses.  The scalar field is left decoupled from ordinary matter in E frame.  This simple situation changes, however, once we take the effect of interaction among matter fields into account.

Consider the quark-gluon interaction, for example.  Calculation based on relativistic quantum field theory suffers always from divergences.  To put this under control we use the recipe of dimensional regularization.  We start with spacetime dimension $N$ which we choose off the physical value 4.  Divergences can be represented by poles $\sim (N-4)^{-1}$ if we confine ourselves to 1-loop diagrams for the moment.  After applying renormalization techniques with $N\neq 4$, we finally go to the limit $N=4$ at the end of the calculation.  We now observe that the scalar field is decoupled from matter only at $N=4$.  In other words, we have the coupling proportional to $N-4$ which may cancel the pole coming from divergences, thus leaving a nonzero finite result.  This is a typical way we obtain ``quantum anomalies,"  which play important roles in particle physics.  In the same way, we are left with nonzero matter coupling of the scalar field in general.

The QCD calculation yields
\beq
{\cal L}_{mq1} = \sqrt{-g_*}\zeta_q m_{q\dagger}\bar{q}_*q_*\sigma,
\label{stu3_21}
\eeq
where 
\beq
\zeta_q =\zeta \frac{5\alpha_s}{\pi}\approx 0.3 \zeta,
\label{stu3_22}
\eeq
with $\alpha_s \approx 0.2$ is the QCD counterpart of the fine-structure constant.   The occurrence of $\alpha_s$ which is specific to the quark implies composition-dependence of the force mediated by $\sigma$, in accordance with violation of WEP, as discussed before.

For more-loop diagrams we obtain terms of higher power of $\sigma$.  Combined with the mass term \reflef{stu3_19}), these coupling terms will give $\sigma$-dependent mass, as before.  We will show, however, that the dependence might be reasonably weaker than in the prototype model.  Consider a nucleon which constitutes a major component of the real world, but is composite to make the same calculation as to the quark not applicable.  In view of a rather small content of the quark mass contribution ($\sim 60$MeV) [\cite{qcont}] out of the total nucleon mass $\sim 940$MeV, we estimate an effective coupling strength of the nucleon;
\beq
\zeta_N \approx 0.02 \zeta.
\label{stu3_23}
\eeq
The smallness of $\zeta_N/\zeta \ll 1$ seems to justify E frame as a physical CF to a good approximation, although strictly speaking we should move to anther CF in which the nucleon mass is truly constant.  By this process $\phi$ reappears in front of scalar curvature, thus making the gravitational constant time-dependent.  The predicted rate of change based on \reflef{stu3_23}) is $\dot{G}/G \sim 10^{-12}{\rm y}^{-1}$, in roughly consistent with the observational upper bound obtained so far [\cite{hel}].

We also find that the exponent of the scale factor in the dust-dominated universe is given by
\beq
\alpha_* =\frac{2}{3}\left( 1-\frac{\bar{\zeta}}{\zeta} \right) \sim \frac{2}{3},
\label{stu3_24}
\eeq
where $\bar{\zeta}$ is an average of $\zeta$'s for particles that comprise dust matter.  In this way we are in a better status compared with the model discussed in Section II.

We also add that scale invariance is broken explicitly finally when the quantum effect is included.  $\sigma$ is now a pseudo NG boson which acquires a nonzero mass.  This seems to raise a tempting possibility that $\sigma$ provides yet another origin of non-Newtonian gravity [\cite{wet}], as will be further pursued.

\section{Non-Newtonian gravity}

We decompose $\sigma$ into the cosmological background part and the locally fluctuating component;
\beq
\sigma(x) = \sigma_b(t_*) + \sigma_f(x).
\label{stu3_25}
\eeq
Substituting this into \reflef{stu3_21}) yields the matter coupling of each component.

Consider the $\sigma_f$-coupling.  A 1-quark-loop diagram will give a self-mass $\mu_f$ basically given by
\beq
\mu_f^2 \sim m_{*q}^2 M_{*sb}^2,
\label{stu3_26}
\eeq
where we assumed the cut-off energy of the order of $M_{*sb}$, the mass scale of supersymmetry breaking.  We use the values $m_{*q}\sim 5$MeV and $M_{*sb}\sim 1$TeV, which are $\sim 2\times 10^{-21}$ and $\sim 4\times 10^{-16}$, respectively in the reduced Planckian unit system, obtaining $\mu_f\sim 0.84\times 10^{-36}\sim 2.1\times 10^{-9}$eV, and the corresponding force-range $\lambda = \mu^{-1}\sim 1.2\times 10^{36} = 9.6 \times 10^3 {\rm cm}\approx 100 {\rm m}$.  This turns out to be a macroscopic size considered typically as the force-range of non-Newtonian gravity [\cite{yfnat}], though obviously latitudes of several orders of magnitude should be allowed.

On the other hand, it has been argued that the force of $\sigma$ should be of long-range [\cite{qt2}].  A simple estimate of this kind can be obtained by considering a component of the Einstein equation $3H_*^2 = \rho_{\sigma}+\rho_{m*}$.  The left-hand side would be $\sim t_*^{-2}$, giving $\rho_{\sigma}\sim V\sim t_*^{-2}$.  We may also expect that the mass-squared $\mu^2$ will be given by $\mu^2 \sim V''\sim V$.  Combining these we reach a conclusion $\lambda \sim t_*$ which is the size of the whole visible universe, in sharp contrast with the previous estimate.

In this connection we point out that the latter argument applies to $\sigma_b$, a classical solution of a {\em nonlinear} equation, while $\mu_f^2$ in the former is a squared  mass of a quantized field, which is based on the solution of a {\em linear} harmonic oscillator.  These two can be simply different from each other.  An example is provided by the sine-Gordon equation in 2 dimensions, featuring a quantized mesonic excitation {\em and} the classical soliton solution, which show behaviors entirely different from each other [\cite{sG}].

We do not know if the 2-dimensional model provides an  appropriate analogy, nor if our cosmological equations in 4 dimensions allow a soliton-type solution.  Nevertheless the lesson learned from this example seems highly suggestive.    Without any further details at this moment, we offer a conjecture that $\sigma_f$ mediates an intermediate-range force between local objects, with $\sigma_b$ still rolling down the exponential slope slowly.

We also point out that $\sigma_b$ may also acquire self-energy through its own matter coupling.  As it turns out, this amounts to be part of the vacuum energy of matter fields.   In the sense of quantum field theory,  this contribution is given roughly by $\sim M_{*sb}^4$ which provides another origin of the cosmological constant with the size represented by $\Omega_{\Lambda}$ as large as $\sim 10^{60}$.  Since $M_{*sb}$ depends on $\sigma$, we would have the potential too large by the same amount.  The same ``success" for the contribution from $\Lambda$ in the J frame Lagrangian dose not seem expected.  An entirely different mechanism based perhaps on nonlinearly or topological nature is called for.  If a solution is discovered finally, however, we would be likely left with non-Newtonian gravity featuring finite force-range and WEP violation.  Quintessence might be quintessentially a fifth force.

The phenomenological aspects of this putative force have been analyzed in terms of the static potential between two nuclei;
\beq
V_{5ab}(r) =-\frac{G m_am_b}{r }\left(1 + \alpha_{5ab} e^{-r/\lambda}\right).
\label{stu3_27}
\eeq
The size of the coefficient $\alpha_{5ab}$ is given essentially by 
\beq
\alpha_{5ab} \sim \alpha_{5N} = 2\zeta_N^2 \sim 10^{-3},
\label{stu3_28}
\eeq
where the estimate at the far right-side is based on \reflef{stu3_23}) together with $\zeta \sim 1$ assumed.  Given the experimental constraints [\cite{fsb}],  we are on the verge of immediate exclusion, though it might be premature to draw a final conclusion before more detailed analyses are completed on various kinds of uncertainties.

\section{Two-scalar model}

We now discuss how the exponential potential should be modified to understand an accelerated universe.  We start with a clue which we find in a {\em transient} solution for the exponential potential.

Suppose $\sigma(t_*) (=\sigma_b)$ starts moving ahead even slightly.  The potential will decrease rapidly, and then the frictional force coming from the cosmological expansion will decelerate $\sigma$ until it comes to an almost complete stop.  $\rho_{\sigma}$ will fall off much faster than $\rho_{m*}$ to reach a near constant.  This state lasts until $\rho_{m*}$ comes down to be comparable again with $\rho_{\sigma}$ when the system enters the asymptotic phase, described by \reflef{stu3_9})-\reflef{stu3_12}).  The plateau of $\rho_{\sigma}$ thus formed is expected to mimic a cosmological constant.

Detailed numerical simulation shows, however, that $\sigma$ starts moving again toward the asymptotic behavior a little too early, before $\rho_{\sigma}$ crosses $\rho_{m*}$.  Remember that this crossing, if it occurred, would have implied $\Omega_{\Lambda}=0.5$.  As a consequence $\Omega_{\Lambda}$ would reach short of values as ``large" as 0.6--0.7.  We must invent a mechanism to prevent this premature departure of $\sigma$.

\begin{figure}[h]
\hspace*{3.5cm}
\epsfxsize=10cm
\epsffile{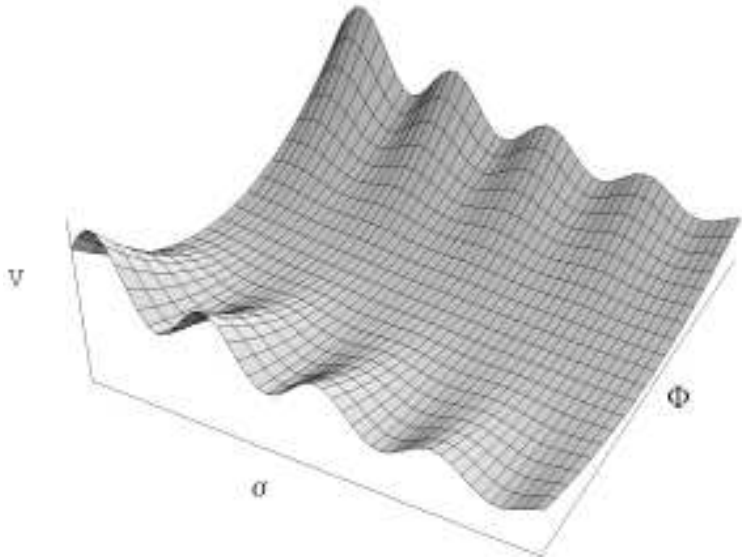}
\caption{The potential $V(\sigma, \Phi)$ given by \protect\reflef{stu3_29}).  Along the central valley with $\Phi =0$, the potential reduces to the simpler behavior $\Lambda e^{-4\zeta\sigma}$ as given by \protect\reflef{stu3_8}), but with $\Phi \neq 0$, it shows an oscillation in the $\sigma$ direction.}
\label{fig1}
\end{figure}

After somewhat random effort on a try-and-error basis, we came across a proposal [\cite{yfplast}].  We introduce another scalar field $\Phi$ which couples to $\sigma$ through the E frame potential,
\beq
V(\sigma, \Phi)=e^{-4\zeta\sigma}\left( \Lambda+\half m^2\Phi^2\left[ 1+\gamma\sin(\kappa\sigma ) \right]\right),
\label{stu3_29}
\eeq
as shown graphically in Fig. \ref{fig1}, where $m, \gamma$ and $\kappa$ are constants roughly of the order unity.  We will study the consequences simply for the success, without any discussion on the theoretical origin, at this moment.  In Fig. \ref{fig2}, we show an example of the solution in which quantities are plotted against $u\equiv \log t_*$.

We notice that $\sigma$ and $\Phi$ in the upper diagram show alternate occurrence of rapid changes and near standstills.  Corresponding to this we find plateau behaviors of $\rho_s$, the energy density of the $\sigma-\Phi$ system, in the lower diagram.  There are two of them in this particular example.  In both of them $\rho_s$ crosses $\rho_{m*}$, resulting in sufficient acceleration of the scale factor, as clearly recognized as two mini-inflations in the upper and middle diagrams.

We moderately fine-tuned initial values and parameters such that the second mini-inflation takes place around the present epoch, $u\approx 60$.  It is rather easy to ``predict" values like $t_{*0}=(1.1-1.4)\times 10^{10}{\rm y}$, $\Omega_{\Lambda} = 0.6-0.8$ and $h = 0.6-0.8$, with more details described in Ref. [\cite{yftwo}].

We provide an intuitive explanation briefly on how we understand this result.  The decelerated $\sigma$ settles at a place which is determined only by the ``initial" values, quite independent of where it is with respect to the sinusoidal potential which had sunk so much.  As the time elapses, the potential grows in terms of the strength relative to the frictional force.  It may happen that $\sigma$ finds itself  near one of the minima of the sinusoidal potential.  $\sigma$ is then trapped, tighter and tighter, as the potential strength proportional to $\Phi^2$ builds up again in the relative sense as above.  This serves to prevent $\sigma$ from escaping too early, and consequently to have the plateau behavior to last sufficiently long.  The growing potential implies, however, that also $\Phi$ is attracted toward the central valley $\Phi = 0$ stronger and stronger.  Finally the time will come when $\Phi$ starts moving downward.  This entails a sudden collapse of the potential wall that has confined $\sigma$.  With the accumulated energy $\sigma$ is catapulted forward.  See Fig. \ref{fig3} showing a magnified view of this behavior.  In short this ``trap-and-catapult mechanism" is a result of a superb collaboration of the two scalar fields.

The decelerated $\sigma$ may find itself again near a potential minima.  Then nearly the same pattern of the behavior will repeat itself, as in the above example.  Otherwise, $\sigma$ and $\Phi$ will make a smooth transition to the asymptotic phase, as in the model without $\Phi$.  No mini-inflation and hence no extra acceleration take place.  There are many different combinations of these typical behaviors.

\section{Discussion}

Our argument in Section IV for a possible link to non-Newtonian gravity will remain unchanged for $\sigma$ even in the two-scalar model.  Experimental searches for the non-Newtonian force with improved accuracy are strongly encouraged.  On the other hand, $\Phi$ will most likely stay extremely massive leaving virtually no trace in the macroscopic physics.

\begin{figure}[h]
\hspace*{2.5cm}
\epsfxsize=11cm
\epsffile{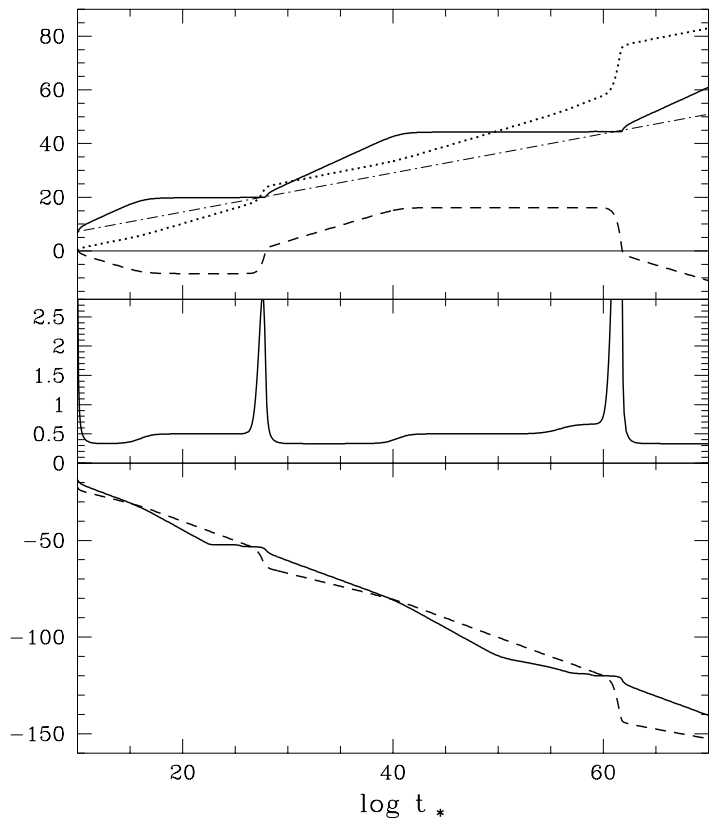}
\caption{An example of the solution.  Upper diagram: $b=\ln a_*$
(dotted), $\sigma$ (solid) and $2 \Phi$ (dashed) are plotted against
$\log t_*$.  The present epoch corresponds to $\log t_* =60.1-60.2$,
while the primordial nucleosynthesis must have taken place at $\log t_* 
\sim 45$.   The parameters are $\Lambda =1, \zeta = 1.5823, m= 4.75, 
\gamma = 0.8, \kappa = 10, \zeta_{\rm d} = 0.002$.  The initial values at $t_{1*} = 10^{10}$
are $\sigma _1=6.7544, \sigma'_1 =0$ (a prime implies differentiation
with respect to $\tau=\ln t_*$),$ \Phi_1 = 0.21, \Phi'_1 = -0.005,
\rho_{1 \rm rad}= 3.7352 \times 10^{-23}, \rho_{1 \rm dust}=4.0 \times
10^{-45}$.  The dashed-dotted straight line represents the asymptote of $\sigma$ given by $\tau /(2\zeta)$.  Notice long plateaus of $\sigma$ and
$\Phi$, and their rapid changes  during relatively ``short''
periods.   Middle diagram: $\alpha_* =b' = t_* H_*$ for an effective
exponent in the local power-law expansion $a_*\sim t_*^{\alpha_*}$ of
the universe.  Notable leveling-offs can be seen for 0.333, 0.5 and 0.667
 corresponding to the epochs dominated by  the kinetic terms of the
scalar fields, the radiation-matter and the dust-matter, respectively.
Lower diagram: $ \log\rho_s$ (solid), the total energy density of the
$\sigma-\Phi$ system, and $\log\rho_{m*}$ (dashed), the matter energy density.
  Notice an interlacing pattern of $\rho_s$ and $\rho_{m*}$, still obeying 
$\sim t_*^{-2}$ as an overall behavior.  Nearly flat plateaus of
$\rho_s$ precede before $\rho_s$ crosses $\rho_{m*}$, hence with
$\Omega_{\Lambda}$ passing through 0.5.
}
\label{fig2}
\end{figure}

\begin{figure}[h]
\hspace*{2.5cm}
\epsfxsize=11cm
\epsffile{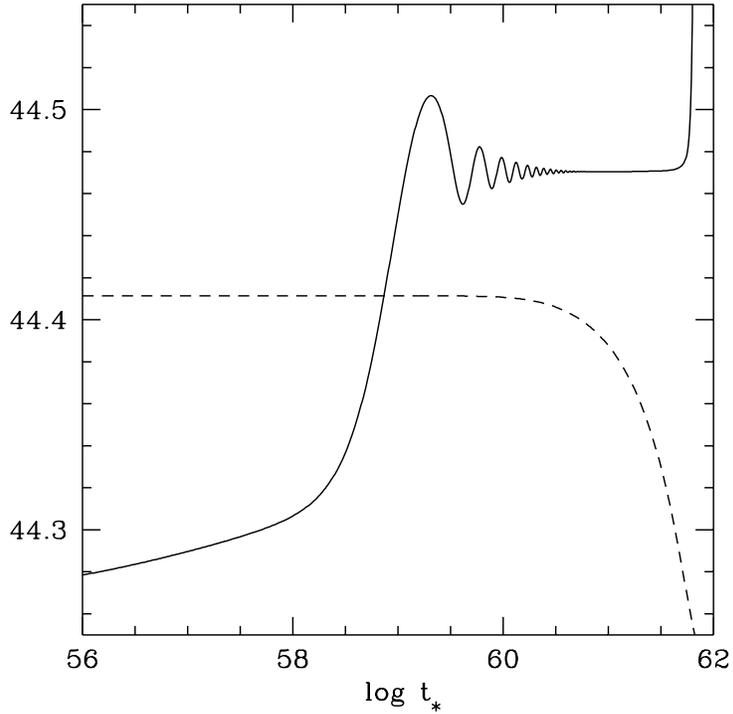}
\caption{Magnified view of $\sigma$ (solid) and $0.02 \Phi+44.25$ (dashed) in the upper diagram of Fig. \protect\ref{fig2}.  Note that the vertical scale has been expanded by approximately 330 times as large compared with Fig. \protect\ref{fig2}.  With the time variable $\tau = \ln t_*$, the potential $V$ grows relatively to the frictional force as the multiplying factor $t_*^2=e^{2\tau}$.  The potential wall for $\sigma$ becomes increasingly steeper, thus confining $\sigma$ further to the bottom of the potential, as noticed by an oscillatory behavior of $\sigma$ with ever increasing frequency measured in $\tau$.  (In this particular example, $\sigma$ was trapped finally long after it had been decelerated to settle near one of the maxima of the potential.)  The growing $V$ causes $\Phi$ eventually to fall downward, resulting in the collapse of the confining potential wall.  The stored energy is then released to unleash $\sigma$.}
\label{fig3}
\end{figure}

In Fig. \ref{fig2} we find $\rho_s$ and $\rho_{m*}$ fall off interlacing each other, maintaining the dependence $\sim t_*^{-2}$ as an overall behavior.  In this way we inherit the scenario of a decaying cosmological constant, unlike many other models of quintessence.

We have not attempted to determine the parameters and initial values uniquely.  We add some remarks, however.  They should be determined such that a plateau behavior has lasted long enough to cover the present epoch and the period of nucleosynthesis, as shown in Fig. \ref{fig2}.  By requiring this we can guarantee that particle masses must have been the same as today, also keeping the period well in the radiation-dominated epoch.  There is a wide area in parameter space where this is not the case, thus endangering one of the most successful part of standard cosmology.

Some of the coupling constants, including $G$, may depend on time
probably through its dependence on scalar fields.  Assuming that we
are today during the period of a plateau behavior entails no time
variability of the coupling constants, or below the level of
$t_{*0}^{-1}\approx 10^{-10}{\rm y}^{-1}$ by perhaps several orders of
magnitude, apparently in good agreement with the upper bounds obtained
so far [\cite{tvar}].  The previous argument on $\dot{G}/G$ in Section
III no longer applies to the epochs during a plateau period.

Given the required long plateau behavior, one may wonder if the modification of the model proposed in Sections II-III loses its ground.  No need is present to distinguish CFs from each other during a plateau behavior.  The argument still holds true, however, for epochs much earlier than nucleosynthesis.  We must avoid concluding a contracting universe, for example, at very early epochs [\cite{yfptp}].

Suppose we are really witnessing an extra acceleration of the universe.  Suppose also this is due to a fixed constant $\Lambda$.  It then follows that we are in a truly remarkable, special once-for-all era throughout the whole history of the universe.  Taking aside a special version of the Anthropic Principle, this should be too arrogant a view, a way of representation of the coincidence problem.  From the point of view of the two-scalar model, on the other hand, we are looking at only  one of the repeated events.  We still face the coincidence problem but certainly to a lesser degree.

We can show that the analytic solution \reflef{stu3_9})-\reflef{stu3_12}) supplemented with $\Phi=0$ is still an asymptotic solution.  Before reaching this fixed-point attractor, however, there can be many different types of transient behaviors, depending sensitively on the initial values.  In particular we have many different possibilities with respect to the number showing how many plateau behaviors we experience before the solution finally enters the asymptotic phase.  This number, and hence the qualitative characteristics of the solution changes suddenly as an initial value changes smoothly.  This reminds us of the well-known chaotic behavior in nonlinear dynamics.  We remind readers that two scalar fields have four degrees of freedom beyond the smallest value three for the system possibly to be chaotic [\cite{corn}].  The presence of the fixed-point attractor rather than the strange attractor in phase space, however, implies that the behavior is not genuinely chaotic, but only chaos-like.  The dependence on initial values and parameters is still so sensitive that the system seems to deserve further scrutiny.  It may provide a new example of dissipative structure.

In this sense we are leaving the traditional view that the universe we see now should depend as little as possible on initial values.  Nothing seems seriously wrong if, on the contrary, today's universe is, like many other phenomena in nature around us, still in a transient state of evolution depending strongly on what the universe was like at very early times [\cite{yfmor}].

We finally add a comment on a possible extension of the
model.\footnote{After completing the manuscript we learned about 
other detailed studies on the extended models [\cite{oth}].}  The starting Lagrangian in J frame may be thought of as the theoretical one obtained by the process of dimensional reduction from the still more fundamental Lagrangian perhaps in higher dimensions.  It is then likely that $\Lambda$ is multiplied by a monomial of $\phi$.  This provides a possible extension of the simplest model considered so far.  As we find, we still reach an exponential potential with $\zeta$ in \reflef{stu3_8}) replaced by another factor depending on the exponent of the monomial.  This suggests that the analyses carried out here can be applied to more general models.

\bigskip
\bcent
{\Large\bf References}
\ecent

\begin{enumerate}
\item\label{perl}For the most recent result, see, for example,  A.G. Riess {\it et al}., Astron. J. {\bf 116} (1998), 1009: S. Perlmutter {\it et al.}, Astroph. J. {\bf 517} (1999), 565.
\item\label{jordan}P. Jordan, Z. Phys. {\bf 157} (1959), 112.
\item\label{bd}C. Brans and R.H. Dicke, Phys. Rev. {\bf 124} (1961), 925.
\item\label{peeb}P.J.E. Peebles and B. Ratra, Astrophys J., {\bf 325} (1988), L17; B. Ratra and P.J.E. Peebles, Phys. Rev. {\bf D37} (1988), 3406; T. Damour and K. Nordtvedt, Phys. Rev. {\bf D48} (1993), 3436; D.I. Santiago, D. Kallingas and R.V. Wagoner, Phys. Rev. {\bf D58} (1998), 124005
\item\label{chiba}T. Chiba, N. Sugiyama and T. Nakamura, Mon. Not. R. Astron. Soc. {\bf 289} (1997), L5.
\item\label{stein}R.R. Caldwell, R. Dave and P.J. Steinhardt, Phys. Rev. Lett. {\bf 80} (1998), 1582. 
\item\label{qt}J.A. Frieman and I. Waga, Phys. Rev. {\bf D57} (1998), 4642; P. Ferreira and M. Joyce, Phys. Rev. {\bf D58} (1998), 023503; A.R. Liddle and R.J. Scherrer, Phys. Rev. {\bf D59} (1999), 023509;  W. Hu, D.J. Eisenstein, M. Tegmark and M. White,  Phys. Rev. {\bf D59} (1999), 023512; G. Huey, L. Wang, R.R. Caldwell and P.J. Steinhardt, Phys. Rev. {\bf D59} (1999), 063005; P.J. Steinhardt, L. Wang and I. Zlatev, Phys. Rev. {\bf D59}(1999), 123504; F. Perrotta, C. Baccigalup and S. Matarrese,  Phys. Rev. {\bf D61} (2000), 023507; M. Bento and O. Bertolami, Gen. Rel. Grav. {\bf 31} (1999), 1461; F. Rosati, hep-ph/9906427.  
\item\label{qt2}S. Carroll, Phys. Rev. Lett. {\bf 81} (1998), 3067; T. Chiba, Phys. Rev. {\bf D60} (1999), 083508; N. Bartolo and M. Pietroni, Phys. Rev. {\bf D61} (2000), 203518.
\item\label{wnb}See, for example, S. Weinberg, Rev. Mod. Phys. {\bf 61}(1989), 1.
\item\label{decay}A.D. Dolgov, {\it The very early universe}, Proceedings of Nuffield Workshop, 1982, Cambridge University Press, 1982; Y. Fujii and T. Nishioka, Phys. Rev. {\bf D42} (1990), 361.
\item\label{yfptp}Y. Fujii, Prog. Theor. Phys. {\bf 99} (1998), 599.
\item\label{yfmor}Y. Fujii, {\it Fundamental parameters in cosmology}, Proceedings of the XXXIIIrd Rencontres de Moriond, Editions Frontieres, 1998, p. 93, gr-qc/9806089.
\item\label{yftwo}Y. Fujii, {\it A two-scalar model for a small but nonzero cosmological constant}, gr-qc/9908021.
\item\label{yfnn}Y. Fujii, {\it Quintessence, scalar-tensor theories and non-Newtonian gravity}, gr-qc/9911064.

\item\label{dpr}R.H. Dicke, Phys. Rev. {\bf 125} (1962), 2163.
\item\label{fsb}E. Fischbach and C.L. Talmadge, {\it The search for non-Newtonian gravity}, AIP Press-Springer, 1998.
\item\label{qcont}M.P. Locher, Nucl. Phys. {\bf A527} (1991), 73c.
\item\label{hel}R.W. Hellings {\it et al}., Phys. Rev. Lett. {\bf 51} (1983), 1609.
\item\label{wet}See also R.D. Peccei, J. Sola and C. Wetterich, Phys. Lett. {\bf B195} (1987), 183.
\item\label{yfnat}Y. Fujii, Nature Phys. Sci. {\bf 234} (1971), 5: Int. J. Mod. Phys. {\bf A6} (1991), 3505.
\item\label{sG}See, for example, S.V. Ketov, Fortsch. Phys. {\bf 45} (1997), 237.
\item\label{yfplast}Y. Fujii and T. Nishioka, Phys. Lett. {\bf B254} (1991), 347; Y. Fujii, Astropart. Phys. {\bf 5} (1996), 133.
\item\label{tvar}See, for example, Y. Fujii, M. Omote and T. Nishioka,
Prog. Theor. Phys. {\bf 92} (1992), 521; Y. Fujii, A. Iwamoto,
T. Fukahori, T. Ohnuki, M. Nakagawa, H. Hidaka, Y. Oura and
P. M\"{o}ller, {\it The nuclear interaction at Oklo 2 billion years
ago}, to be published in Nucl. Phys. {\bf B}, hep-ph/9809549v2, and papers cited therein.
\item\label{corn}See, for example, J.N. Cornish and J. Levin,
Phys. Rev. {\bf D53} (1996), 3022; R. Easther and K. Maeda,
Class. Quant. Grav. {\bf 16} (1999), 1637.
\item\label{oth}L. Amendola, Phys. Rev. {\bf D60}(1999), 043501;  J.P. Uzan, Phys. Rev. {\bf D59}(1999), 123510; F. Perrotta, C. Baccigalup and S. Matarrese, Phys. Rev. {\bf D61}(2000), 023507; O. Bertolami and P.J. Martins, gr-qc/9910056.
\end{enumerate}
\mbox{}\\[5em]
\mbox{}\\[2em]

\end{document}